\def\beg{\begin{equation}}
\def\eeq{\end{equation}}
\documentstyle[12pt]{article}
\begin{document}
\begin{center}
{\Large{\bf Spin-dependent charge of quasiparticle 
clusters 
in quantum Hall effect:Interpretation of unsolved data of Pan, 
Stormer, Tsui, Pfeiffer, Baldwin and West.}}
\vskip0.35cm
{\bf Keshav N. Shrivastava}
\vskip0.25cm
{\it School of Physics, University of Hyderabad,\\
Hyderabad  500046, India}
\end{center}

A state with 2 electons has a charge of 1/2. The states of 3 
electrons have quasiparticle charge of 4/11 and 5/13; the one 
with 5 electrons have quasiparticle charges 4/13 and 6/17. The 
state with 7 electrons exhibits quasiparticle charges of 5/17 
and 12/17. 
When spin is reversed, the charge of three particle state changes 
from 4/11 to 7/11. These predicted results agree with the recent 
experimental measurements of Pan, Stormer, Tsui et al. In addition, 
we predict a state of charge 3/11, not given in the data.
\vskip1.0cm
Corresponding author: keshav@mailaps.org\\
Fax: +91-40-2301 0145.Phone: 2301 0811.
\vskip1.0cm

\noindent {\bf 1.~ Introduction}

     Recently, Pan et al [1] have obtained the effective fractional 
charges of $\nu$= 4/11, 5/13, 6/17, 4/13, 5/17 and 7/11, the origin 
of which is unsolved. Last year, we have written a book which 
explains the quantum Hall effect [2]. All of the predicted fractions 
are the same as those found in the experimental measurements of 
Stormer [3]. Therefore,
we tried to understand these fractional charges which are given 
above. Earlier, the effort was limited to spin 1/2 only. We now 
find that if
 we take clusters of quasiparticles, we can generate more values of 
the fractional charge. When 3, 6 or 8 quasiparticles are taken all 
of the fractional charges are predicted correctly.

     In this paper, we describe the relationship between spin and 
charge and calculate the quasiparticle charge. When spin is reversed, 
the charge 4/11 changes to 7/11. The 4/11 and the 7/11 are thus Kramers
 time reversed states.

\noindent{\bf 2.~~Fractional charge}

     We have found that the quasiparticle charge depends on spin, s, 
as,
\beg
e_{eff}/e={{\it l}+{1\over 2} \pm s\over 2{\it l}+1}
\eeq
which has the odd denominator. If a half fraction comes from the 
numerator, the denominator will cease to be odd. When ${\it l}$=0, 
$s$=1/2, the above gives the effective charge of,
\beg
e_{eff}=0\,\,\, or \,\,\, 1.
\eeq
Therefore, for {\it l}=0, the quasiparticles can be electrically 
neutral
or they have the charge equal to that of the electron. When clusters 
of electrons are formed, the quasiparticle spin need not be 1/2.
 Hence, additional values of spin have to be taken into account 
which give a variety of charges.

     Let us consider $s$=1/2 but two particles. When the pair is 
a singlet, $S$=0 means that one particle has spin +1/2 and the 
other spin -1/2, so that the formula (1) for ${\it l}$=0, $s$=0
 gives,
\beg
e_{eff}/e = {1\over 2}.
\eeq
This is strange that ${\it l}$=0, $s$=0 state has a quasiparticle
 of charge 1/2. In fact, this value of 1/2 although obtained here 
from nonrelativistic theory, agrees with Jackiw and Rebbi [4] who 
obtained 1/2 of a fermion in the $1+1$ dimensions in a relativistic 
theory. In the experimental data on quantum Hall effect, 1/2 an 
effective charge 
is quite often observed. The state ${\it l}$=0  and $s$=0, should 
be
superconducting but in the present case, as the magnetic field is 
varied, the state is destroyed. For $s$=1, we have spin triplets. 
When we substitute ${\it l}$=1, 2 or 3 with $s$ =1, we obtain the 
fractional
charge $\nu_+$ with spin +1 and $\nu_-$ with $s$=-1 as given below.
\begin{center}
\begin{tabular}{cccc}
\hline
{\it l}& $s$ & $\nu_+$ & $\nu_-$\\
\hline
1&1&5/6&1/6\\
2&1&7/10&3/10\\
3&1&9/14&5/14\\
\hline
\end{tabular}
\end{center}
\vskip0.25cm
which have even denominators. For $s$=3/2, we get,
\begin{center}
\begin{tabular}{cccc}
\hline
{\it l}& $s$ & $\nu_+$ & $\nu_-$\\
\hline
1 & 3/2& 1 & 0\\
2 & 3/2& 4/5 & 1/5\\
3 & 3/2& 5/7 & 2/7\\
\hline
\end{tabular}
\end{center}
\vskip0.25cm
which has a normal solution with effective charge 1 and a 
charge-density wave type solution with charge $0$ and one other 
solution, such as 1/5.
Here, it will be {\it sufficient}, if we can follow those fractions 
which have been left {\it unsolved} by Pan et al[1]. Therefore, 
we tabulate ${\it l}$, $s$ and $\nu_+$ and $\nu_-$ for three, five 
and seven electrons. The spin is written according to Dirac's result
 so 
that the electrons have parallel spin due to Coulomb interaction.
 For 
3 electrons the spin is 3/2, for five electrons it is 5/2 and for
 7 electrons it is 7/2. We take ${\it l}$=5, 6 and 8. For positive
 sign 
the value of (1) is called $\nu_+$ and for negative sign in (1) we 
have $\nu_-$. Therefore the values are as follows:
\begin{center}
\begin{tabular}{cccc}
\hline
{\it l}& $s$ & $\nu_+$ & $\nu_-$\\
\hline
5 & 3/2 & 7/11 & 4/11\\
5 & 5/2 & 8/11 & 3/11\\
6 & 3/2 & 8/13 & 5/13\\
6 & 5/2 & 9/13 & 4/13\\
8 & 5/2 & 11/17 & 6/17\\
8 & 7/2 & 12/17 & 5/17\\
\hline
\end{tabular}
\end{center}
\skip0.25cm
The $\nu_-$ have negative sign in eq.(1) and $\nu_+$ have positive 
sign.
Thus the values of  4/11, 5/13, 4/13, 6/17 and 5/17 belong to 
clusters of quasiparticles, all of which have negative sign and 
hence spin polarized. The fraction 7/11 has positive sign and 
hence its polarization is opposite to that of 4/11. Indeed, as the 
magnetic 
field is rotated, 7/11 becomes dim or vanishes. This means that the 
spins 
tilt when field is rotated. We find that the value ${\it l}$=7 is 
missing in the above table. That is because ${\it l}$=7 gives 9/15 
and 6/15 which are the same as 3/5 and 2/5 already known for
 ${\it l}$=2. What has happened is that the denominator
 2${\it l}$+1 becomes factorizable and the effect of spin is to 
cancell part of the value arising from 2${\it l}$+1. For three
 quasiparticles $s$=3/2 and ${\it l}$=7, the same charge is found 
as for $s$=1/2, ${\it l}$=2. Thus the effect of the spin is to 
reduce the orbital value. Pan et al[1] find that these sequences 
do not fit into the standard series. On the other hand, it is clear 
that our formula (1) works better than that of Pan 
et al and obviously agrees with the experimental data. In the table
 above, we have included ${\it l}$=5 and $s$=5/2. Therefore, we
 predict 3/11 for the negative sign and 8/11 for the positive sign. 
The 3/11 not yet found in the figure of Pan et al is predicted before 
observation. 
It requires slightly higher field than is plotted.

\noindent{\bf3.~~ Discussions}.

     We have previously predicted all of the Stormer's fractional 
charges correctly by using $s$=$\pm$ 1/2. In the present note we 
realize that clusters of quasiparticles are formed so that the spin 
can be 1, 3/2, 5/2, 7/2, etc. When these values of spin are used 
the fractional charges of 4/11, 5/13, 4/13, 6/17 and 5/17 are 
predicted in agreement with the experimental data of Pan et al. 
We also predict 
3/11 which is not found in the experimental data.

\noindent{\bf4.~~Conclusions}.

     In conclusion, we find that there are clusters of the same 
spin as predicted by Dirac's results of particles in a Coulomb 
potential. There are effective charges  due to a new spin-charge  
relationship.

\vskip1.25cm

\noindent{\bf5.~~References}
\begin{enumerate}
\item W. Pan, H. L. Stormer, D. C. Tsui, L. N. Pfeiffer, K. W. 
Baldwin and K. W. West, Phys. Rev. Lett.{\bf 90}, 016801 (2003).
\item K. N. Shrivastava, Introduction to quantum Hall effect, 
Nova Science, New York, N. Y. (2002).
\item H. L. Stormer, Rev. Mod. Phys. {\bf 71 }, 875 (1999).
\item R. Jackiw and C. Rebbi, Phys. Rev. D {\bf13 }, 3398 (1976)
\end{enumerate}
\vskip0.1cm

Note: Ref.2 is available from:
 Nova Science Publishers, Inc.,
400 Oser Avenue, Suite 1600,
 Hauppauge, N. Y.. 11788-3619,
Tel.(631)-231-7269, Fax: (631)-231-8175,
 ISBN 1-59033-419-1 US$\$69$.
E-mail: novascience@Earthlink.net

\end{document}